\documentclass[twoside]{article}
\usepackage{fleqn,espcrc2}
\usepackage{graphicx}
\usepackage{here}

\newcommand{\AmS}{{\protect\the\textfont2
    A\kern-.1667em\lower.5ex\hbox{M}\kern-.125emS}}
\def\beq{\begin{equation}}
\def\eeq{\end{equation}}
\def\bea{\begin{eqnarray}}
\def\eea{\end{eqnarray}}
\def\bq{\begin{quote}}
\def\eq{\end{quote}}

\def\nnb{\nonumber}
\def\ga{\left(}
\def\dr{\right)}

\def\rar{\rightarrow}
\def\lrar{\Longrightarrow}

\def\nnb{\nonumber}
\def\la{\langle}
\def\ra{\rangle}
\def\nin{\noindent}
\def\ba{\begin{array}}
\def\ea{\end{array}}

\def\b{$\bullet$}
\def\als{\alpha_s}

\def\gg2{ \la\alpha_s G^2 \ra}
\def\gg3{g^3f_{abc}\la G^aG^bG^c \ra}
\def\ggg4{\la\als^2G^4\ra}

\title{\bf{\boldmath
{\Large The SVZ-Expansion and Beyond} }\thanks{Talk given at the Workshop Understanding Confinement (Ringberg
Castle,  16-21th May 2005) and at QCD 05 International Conference (Montpellier, 4-8th July 2005)}}
\author{
Stephan Narison\address{ Laboratoire
de Physique Th\'eorique et Astroparticules, Universit\'e
de Montpellier II, Case 070, Place Eug\`ene
Bataillon, 34095 - Montpellier Cedex 05,
France.\\ E-mail:
snarison@yahoo.fr}
 }
\begin{document}
\textwidth 16.5cm
\textheight 21.2cm
\evensidemargin -0.2cm
\pagestyle{empty}
\pagestyle{plain}
\begin{abstract}
\noindent
I discuss the standard SVZ-expansion of the QCD two-point
correlators in terms of the QCD vacuum condensates and beyond it due to a new unflavoured
$1/Q^2$-term (tachyonic gluon mass squared $\lambda^2$) which modelizes the effects of
uncalculated higher order perturbative terms. The approach is confronted with low-energy
and lattice data. One can notice that, in the different examples studied here, high-dimension
condensates are expected to deviate largely from their vacuum
saturation (large $N_c$) values, while the new
$1/Q^2$-term due to $\lambda^2$ solves the
hadronic mass scale hierarchy-puzzle encountered in the early SVZ-sum rule analysis and improves the low-energy
phenomenology. From tau-decay data, one can extract the running strange quark mass $\overline m_s(2~\rm
GeV)=(93^{+29}_{-32})$ MeV and a more accurate value of the QCD coupling $\alpha_s(M_Z)=0.117\pm 0.002$.
\end{abstract}
\maketitle
\section{Introduction}
\nin
The SVZ approach \cite{SVZ} (for a review, see e.g. \cite{SNB}), where the
non-perturbative effects are approximated by the contributions of QCD condensates, continues
to explain successfully the low-energy hadron phenomenology.  Among the different predictions,
one has been able to
predict the $\rho$-meson mass and coupling by the introduction of the gluon condensate
$\la\alpha_s G^2\ra$ \cite{SVZ}, while more recently, one has 
extracted with a high accuracy (after runned until $M_Z$) the QCD coupling
$\alpha_s$ using tau-decay data \cite{BNP}. However, in the beginning, most of QCD perturbative practioners,
have not been enthousiastic on these successes due to the relative small mass scale
at which, one a priori, (na\"\i vely) expects pQCD to breakdown.  The SVZ approach has also been used
as a serious alternative and guide to the lattice calculations, both approaches being based
on the same uses of the QCD fundamental Lagrangian and related parameters. However, unlike lattice QCD, the
accuracy of the SVZ-sum rule is limited and cannot (a priori) be iteratively improved being an
approximate scheme. One of the limitation of the approach is the absence of our knowledge of the complete
pQCD series. The aim of this talk is, firstly, to review the different values of the QCD condensates extracted
from low-energy and lattice data, and, secondly, to present a string-inspired model (tachyonic gluon mass)
\cite{ZAK,CNZ} inducing a new $1/Q^2$-term in the OPE which can mimic the unknown 
higher order short-distance terms of the pQCD series.  Consequences of the model on the well-established
hadron phenomenology \cite{CNZ,SNQ2,SNV}, on the determination of the strange quark mass from tau-decays
\cite{SNMS} will be reviewed. New result on the effect on the accurate determination of  
$\alpha_s$ from tau decays will be
presented.
\section{SVZ-expansion and QCD condensates}
\nin
For a pedagogical introduction, let's consider the
 generic two-point hadronic correlator:
\bea
 &\Pi(q^2)= i\int d^4 x~ e^{iqx}\la 0|{\cal T} J(x)\ga
J(0)\dr^{\dagger}|0\ra
\eea
 built from the local hadronic current
$
J(x).
$
Following SVZ, the correlator can be approximated by a sum of power
corrections:
\beq
 \Pi(Q^2)\simeq\sum_{d\geq 2}{C_{2d}\la{\cal O}_{2d}\ra\over (-q^2)^d}
\eeq
where: $\la{\cal O}_{2d}\ra$ are the
QCD non-perturbative condensates  of dimension
 $D\equiv 2d$;
$C_{2d}$ is the associated perturbative Wilson coefficient;
$q^2\equiv -(Q^2 >0)$ 
is the momentum transfer. Owing to  gauge invariance, there is no $D=2$ term in the chiral limit $m_j=0$, because
the possible $D=2$ term $\la A_\mu A^\mu\ra$ is a priori gauge dependent (see, however
\cite{STODO}). The well-known condensate is the chiral condensate: $\la\bar\psi\psi\ra$ entering into the GMOR-PCAC
relation:
 $(m_u+m_d)\la\bar\psi\psi\ra=-f_\pi^2m_\pi^2: f_\pi=93 MeV.$ We show in Table \ref{tab:condensate} the ratio
of the {\it normal ordered} condensates $\la\bar ss\ra/\la\bar dd\ra$, indicating large SU(3) breakings. \\
Less-known condensates are the SVZ-condensates entering into the OPE:\\
\b~The Gluon Condensate $\la\alpha_s G^a_{\mu\nu}G_a^{\mu\nu}\ra,$
initially introduced by SVZ \cite{SVZ} and estimated phenomenologically by
different groups in the literature  \cite{SVZ,SNB,YNDU,IOFFE05,REINDERS,BELL,BERTL,TARRACH,PAPA,SNG}
including lattice calculations
\cite{GIACO,MARCHESINI,RAKOW} with a spread of results. However,  its value is found to be definitely non-zero and positive. Some of
the results including the SVZ value do not satisfy the lower bound derived by Bell and Bertlmann \cite{BELL}
from moment sum rules. Recent estimates in \cite{SNG,SNB} satisfying this bound from light and heavy
quark channels are given in Table \ref{tab:condensate}.
\\ 
\b~The Mixed Quark-Gluon condensate
$g\la\bar\psi G^{\mu\nu}_a{\lambda_a\over
2}\sigma_{\mu\nu}\psi\ra$ $\simeq M^2_0\la\bar\psi\psi\ra,$
where $M_0^2$ has been estimated from baryon sum rules
\cite{DOSCH,JAMIN}, the $B-B^*$ mass-splittings \cite{SNHEAVY}, string
model \cite{SIMONOV} and lattice calculations, with a fair agreement.\\
\b~The Four-quark Condensates $\la\bar\psi\Gamma_1\psi\bar\psi\Gamma_2\psi\ra$ estimated from different channels
\cite{TARRACH,SNB} and lead to a violation of the factorization assumption (large $N_c$ approximation) by
at least a factor 2.\\
\b~The triple gluon condensate $\la g^3f_{abc}G^aG^bG^c\ra ,...$
originally estimated by SVZ using instanton liquid model
and confirmed later on by lattice calculations \cite{GIACO}.\\
\b~The dimension $D=8$ and higher dimension condensates
estimated in the $V$ \cite{SNG}, $V+A$ and $P+S$ \cite{SNV} channels using
experimental and lattice data \cite{DEGRAND}.\\
\b~More recently, condensates of the $V-A$ channel have been
estimated from $\tau$-decay data \cite{SNV-A} and from QCD at large $N_c$ \cite{FRIOT}.  Large
violations of the vacuum saturation estimates of these condensates have been also observed. 
\\
The results from the different estimates of these condensates are
summarized in Table \ref{tab:condensate}. Tests of these results from other methods such as lattice with dynamical
quarks are mandatory.
\begin{table*}[hbt]
\setlength{\tabcolsep}{1.5pc}
\newlength{\digitwidth} \settowidth{\digitwidth}{\rm 0}
\catcode`?=\active \def?{\kern\digitwidth}
\caption{Values of the ``Standard" QCD condensates compiled in \cite{SNB}.}
\label{tab:condensate}
\begin{tabular*}{\textwidth}{@{}l@{\extracolsep{\fill}}lll}
\\ ``Standard" Condensates & Dimension D&  Values [GeV]$^D$\\
\hline
 $\la\bar ss\ra/\la\bar uu\ra$ (normal ordered) \footnote{...}&3--3=0& $0.66\pm 0.10$\\
$\la\alpha_s G^2\ra$& 4& $(7.1\pm 0.9)\times 10^{-2}$\\
$g\la\bar\psi G^{\mu\nu}_a{\lambda_a\over 2}\sigma_{\mu\nu}\psi\ra\equiv M^2_0\la\bar \psi\psi\ra$& 5&
$M^2_0=(0.8\pm 0.1)~ {\rm GeV} ^2$\\
$\la g^3f_{abc}G^aG^bG^c\ra$&6&$\approx 1.2 ~{\rm GeV} ^2\la\alpha_s G^2\ra$\\
$\la\bar\psi\Gamma_1\psi\bar\psi\Gamma_2\psi\ra$&6& violation of factorization by 2--3\\
$\la GGGG\ra$&8&  violation of factorization by $ \geq 4$\\
\hline
\end{tabular*}
{\footnotesize 
\begin{quote}
$^{1}\,$ Fo the non-normal ordered condensate which possesses a small perturbative piece, the ratio is $0.75\pm 0.12$.
\end{quote}}
\end{table*}
\nin
\section{\boldmath $1/Q^2$-term beyond the SVZ-expansion}
\begin{table*}[hbt]
\setlength{\tabcolsep}{1.5pc}
\catcode`?=\active \def?{\kern\digitwidth}
\caption{$\lambda^2$ from the pion channel}
\label{tab:lambda2}
\begin{tabular*}{\textwidth}{@{}l@{\extracolsep{\fill}}ccccc}
\hline
{{$\tau$}{[GeV$^{-2}$]}}&${\cal
L}_{exp}/ (2f_\pi^2m_\pi^4)$&$R_{exp}$&
$ R^{\lambda^2=0}_{QCD}$&$-a_s\lambda^2$[GeV]$^2$\\ 
\hline
 1.4&1.20&$0.27\pm
0.06$&$0.66^{+0.50}_{-0.31}$&$0.09^{+0.08}_{-0.06}$ \\
1.2&1.29&$0.36\pm
0.07$&$0.79^{+0.42}_{-0.28}$&$0.10^{+0.09}_{-0.05}$\\ 
1.0&1.42&$0.46\pm 0.10$&$0.94^{+0.34}_{-0.25}$&$0.11^{+0.07}_{-0.06}$\\
 0.8&1.61&$0.58\pm
0.12$&$1.10^{+0.26}_{-0.21}$&$0.12\pm 0.06$\\ 
0.6&1.89&$0.73\pm
0.15$&$1.24^{+0.20}_{-0.17}$&$0.12\pm{0.06}$\\ 
 0.4&2.29&$0.89\pm
0.16$&$1.33^{+0.15}_{-0.13}$&$0.10\pm 0.06$\\ 
\hline 
\end{tabular*}
\end{table*}
\nin
A possible existence of this $1/Q^2$-term  as an eventual source of errors in the determination of $\alpha_s$
from tau-decay data has been advocated in
\cite{ALTARELLI}. This conjecture has been checked from a fit of the
$e^+e^-\rar$  hadrons data  \cite{SNQ2} which allow the existence of a small but {\it imaginary}
constituent quark mass:
\beq
m^2_q=-(71\sim 114)^2~{\rm MeV}^2~,
\eeq
after the identification of the result with the quark mass correction of $-6m^2/Q^2$ in this channel. Later on,
it has been found more convenient to identify this term with the tachyonic gluon mass squared $\lambda^2$ \cite{CNZ}.\\
One possible origin of this term might be the $\la A_\mu A^\mu\ra$ condensate which has been found to be
non-zero \cite{ZAK,PEPE}, but gauge dependent. However, there are some proposals that its minimal value is gauge
invariant over the gauge orbit \cite{STODO}. In this talk, I shall consider that the $1/Q^2$-term is
purely of short distance nature and can mainly emerge from the resummation of the infinite terms of the
pQCD series (UV renormalon). From the form of the Cornell potential, one can relate the string tension to
the squared gluon mass $\lambda^2$ \cite{ZAK}:
\beq
V(r)=-{4\over 3}{\alpha_s \over r}+\sigma r \lrar \sigma\approx-{2\over 3}{\alpha_s}\lambda^2
\eeq
giving $\lambda^2\approx -(0.5\sim 1.0)~{\rm GeV}^2.$ A possible indication of the existence of this term
comes from the lattice calculation of the plaquette $\la G^2\ra$, which can be written as:
\beq
\la G^2\ra\simeq {(N_c^2-1)\over a^4}\Bigg{[}\sum_{0}^{n}c'_n\alpha_s^n+ {d'_n\lambda^2a^2
} + c_4\Lambda^4 \Bigg{]} 
\eeq
where $a$ is the lattice spacing; $\Lambda$ is the QCD scale; $c'_n,~ d'_n$ are log- or numerical
coefficients. In the present model,
$d'_n$ is expected to decrease for increasing $n$. 
In fact, for $n\leq 11$, the fit of the lattice data requires a quadratic term \cite{MARCHESINI}, while for
large $n\geq 26$ this term disappears \cite{RAKOW}. In the analytic approach where the pQCD series is at
best known to ${\cal O}(\alpha_s^4)$ \cite{KOSTJA}, we then expect that $\lambda^2$ plays an essential
role for quantifying the unknown higher order terms. In practice, one can systematically introduce
the UV
$\lambda^2$-effect by replacing the gluon propagator at short distance~\cite{CNZ}:
\beq
D^{\mu\nu}(k^2)={\delta^{\mu\nu}\over k^2}\rar \delta^{\mu\nu}\ga {1\over k^2}+{\lambda^2\over k^4}\dr~,
\eeq
where, this operation is gauge invariant to leading order \footnote{A generalization of this procedure to
the whole range of $k^2$ has been also proposed by 't Hooft \cite{tHOOFT} in an exploratory analytic approach
to confinement.}.
\section{Estimate of the tachyonic gluon mass squared  \boldmath $\lambda^2$}
\subsection*{\b~From \boldmath $ e^+e^-\rar$ hadrons data}
\nin
The contribution of the tachyonic gluon mass squared $\lambda^2$ to the vector correlator is \cite{CNZ,BENEKE}:
\beq \Pi_{\rho}(M^2) =
\Pi^0
\Bigg{[}
1+
\left(\frac{{\alpha_s}}{\pi}\right)
\left(1 -1.05 \frac{\lambda^2}{M^2}
\right)
\Bigg{]}
\label{Bor.Pi1.male}
{}
\eeq
Using a ratio of exponential moments (less sensitive to $\alpha_s$) :
\beq {\cal R}\equiv -{d\over d\tau}\log{\int_0^\infty dt ~{\rm e}^{-t\tau}~\frac{1}{\pi}{\rm Im}\Pi}(t)
\eeq
$(\tau\equiv 1/M^2$ sum rule variable), an earlier fit of the data for a
postulate existence of a
 ``$1/M^2$-term"
interpreted in term of $\lambda^2$ leads to \cite{SNQ2}:
\bea
\lambda^2 (1~\rm{GeV})&\approx& -(0.2-0.5)~\rm{GeV}^2~\lrar\nnb\\ 
a_s\lambda^2 &\approx& -(0.03-0.09)~\rm{GeV}^2~,
\eea
with: $a_s\equiv (\alpha_s/\pi)(1~\rm{GeV})\simeq 0.17\pm 0.02$.
\subsection*{\b~From the pion channel}
\nin
 We use again the ratio of moments ${\cal R}$ in order to eliminate $\alpha_s$ and quark mass terms to leading order. We also
use the Pion pole + ChPT parametrization of $3\pi$ contributions+QCD continuum of the
spectral function \cite{BIJNENS}. Then, we deduce the value of $\lambda^2$ in Table \ref{tab:lambda2}. One can notice that
the introduction of $\lambda^2$ has enlarged the stability/QCD duality region to larger $\tau$ (lower $M^2$)
-values which reduces the distance from the pion pole, which is a success of the approach. Taking the average of the previous
results from $e^+e^-$ and $\pi$-channel leads to:
\bea\label{lambda2}
&&(\alpha_s/\pi)\lambda^2\simeq
-(0.07\pm 0.03)~\rm{GeV}^2~~~~\Longrightarrow\nnb\\
&&\lambda^2 (\tau\approx
1~\rm{GeV}^{-2}) \simeq -(0.41\pm 0.18)~\rm{GeV}^2~
\eea
\section{Hadronic scale puzzle and \boldmath $\lambda^2$}
\nin
In the sum rules with $\lambda^2=0$, one has the optimization scale hierarchy :
\bea M_{\rho-canal}\ll M_{\pi-canal}\ll M_{gluonium-canal}.\eea
There are various ways to get these results \cite{NSVZ,CNZ}:
\b~Detailed sum rules analysis or/and 10\% correction criterion leads to:
\bea M^2_{\rho-canal} &\simeq& (0.6\sim 0.8)~{\rm GeV}^2\nnb\\
&\approx& \ga 10{\pi\over 3}
\alpha_s G^2\dr^{1/2},
\eea
indicating the breaking of Asymptotic Freedom (AF) by infrared phenomena !\\
 \b~Using the positivity of the pseudoscalar spectral function, one can deduce:
\bea\label{pion1}
    M^2_{\pi-canal}&\geq&\sqrt{{16\pi^2\over 3}
{f^2_{\pi}m_{\pi}^4\over (\bar{m}_u+\bar{m}_d)^2}}\nnb\\
&\approx& 1.8~\rm{GeV}^2~,
\eea
  where the factor $m_\pi/(mu+md)\approx 13$  brings a large numerical factor!.
 We have seen in the previous analysis that the presence of $\lambda^2$ leads to \footnote{For an alternative
explanation using diquarks (see e.g. \cite{SV}).}:
\beq
    M^2_{\pi-canal}\approx M^2_{\rho-canal}\approx 1~{\rm GeV}^2~.
\eeq
\b~The $I=1$ $a_0$ scalar meson channel has the same scale as the pseudoscalar one because the 
alone difference comes from the $D=6$ condensate contribution which is a small
correction. Contrary to the case of instanton, the $\lambda^2$ contribution
is not affected by chirality which is the main difference between the two approaches that can be checked by
accurate lattice measurements. One has to remind that the sum rule with $\lambda^2=0$ reproduces quite well the
$a_0$ phenomenolgy \cite{SNB,SNS}\\
   \b~In the scalar glueball channel, the uses of the substracted sum rule plus a 
low-energy theorem   (LET)    for $\Pi_G(0)$ give:
\beq
\Pi_G(M^2)\approx
\Pi_G^0\cdot\Bigg{[}1+\left(\frac{8\pi}{-\beta_1}\right)
\left(\frac{\pi}{\alpha_s}\right)^2
\frac{ \langle\alpha_s G^2 \rangle}{M^4}\Bigg{]}, \label{huge}
\eeq
 { where $-\beta_1$=9/2 for 3 flavours, and LET brings a huge factor of about 400 in front of $1/M^4$
{     
\cite{NSVZ}}:
   \beq\label{glue2}
M^2_{0^+-gluonium}\approx 20M^2_{\rho-canal}
\approx~15~\rm{GeV}^2~,
\eeq
which indicates that it is  difficult to interpret the breaking of AF in terms of resonances!   The
inclusion of the
$\lambda^2$ term leads to:
\beq
\label{GG_scalar_res}
\Pi_G(M^2) = \Pi_G^0
\left(1-{3\lambda^2\over M^2}+...\right) {}. \label{scalarGG:res}~.
\eeq
Thus, the $\lambda^2$ correction is expected to be large  due to the absence of an extra power of $\alpha_s$.    A 10\%
correction like in the $\rho$-case gives:
\beq
\label{mcglue}M^2_{0^+-gluonium}~\approx~13 ~\rm{GeV}^2~, 
\eeq
in amusing agreement with the independent estimate in Eq. (\ref{glue2}). More detailed phenomenology of the (pseudo)scalar
gluonia channels lead to a lower value \cite{SNGLUE,SHORE}:
\beq
    M^2_{0^+-gluonium}~\approx~(3\sim 5) ~\rm{GeV}^2 
\eeq
which is still much larger than $M^2_\rho$ !}, indicating the particularity of this channel. Also contrary to the instanton
approach, where the $2^{++}$ channel is not affected by instanton \cite{SHURYAK}, the contribution of $\lambda^2$ is similar to the
(pseudo)scalar channels indicating an universality of the gluonia scales. Large scale is also expected for the hybrid channel as $\lambda^2$ does not also have an extra power of $\alpha_s$ \cite{CN}.

\begin{figure}[htb]
\includegraphics[width=7cm]{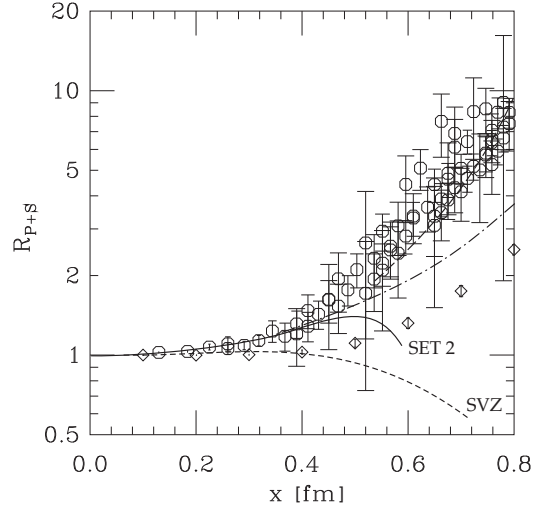}
\caption{Lattice data of the S+P correlator: the dotted curve is the SVZ value (SET~1); the dot-dashed
curve is the prediction for SET~3 where the contribution of the $x^2$-term has been added to SET~2; the
bold dashed curve is SET~3 + a fitted value of the $D=8$ condensate contributions ($\approx (x/0.58)^8$
compared~ to
$(x/1.2)^8$ from factorization); the diamond curve is the prediction from the instanton liquid
model of \cite{SHURYAK}.}
\label{figure: 7}
\end{figure}
\vspace{-1.cm}
\nin
\begin{table}[hbt]
\setlength{\tabcolsep}{.5pc}
\catcode`?=\active \def?{\kern\digitwidth}
\caption{Different sets of the power corrections values}
\label{tab:powercorr}
\begin{tabular*}{7cm}{@{}l@{\extracolsep{\fill}}lccc}
\hline
 Sources&$\la\alpha_s G^2\ra$&$\alpha_s\la\bar\psi\psi\ra
^2$&$(\alpha_s/\pi)\lambda^2$\\
\hline
SET 1\cite{SVZ}
&0.04& $0.25^6$&0\\
SET 2 \cite{SNG}&0.07&$5.8\times
10^{-4}$&$0$\\
SET 3\cite{CNZ,SNQ2}&0.07&$5.8\times
10^{-4}$&$-0.12$\\
\hline
\end{tabular*}
\end{table}
\nin
\section{{\boldmath$ \lambda^2$} versus lattice data and instanton liquid model} 
\nin
For an illustration of the previous discussions both for the
``standard condensates" and for the
$\lambda^2$-contribution, we consider explicitly the scalar+pseudoscalar correlators
$S+P$  in
$x$-space \footnote{Similar data in the $V+A$ channel are also available from the lattice, but do not involve the
$\lambda^2$-contribution because the lattice measures the trace over Lorentz indices which corresponds to
$Q^2\Pi(Q^2)$. The analysis can be found in \cite{SNV}.} where the single instanton effect cancels out:
\bea\label{splusp}
R_{P+S}&\equiv& \frac{1}{2}\ga{\Pi^{P}\over
\Pi^{P}_{pert}}+{\Pi^{S}\over \Pi^{S}_{pert}}\dr
~\rightarrow~
1-{\alpha_s\over
2\pi}\lambda^2x^2\nnb\\
&+&{\pi\over
96}\la\alpha_s(G_{\mu\nu}^a)^2\ra
x^4+{4\pi^3\over
81}\alpha_s\la\bar{q}q\ra^2x^6\ln
x^2\nnb\\
\eea 
Lattice data on $R_{P+S}$ are available \cite{DEGRAND} and we show the analysis in Fig. \ref{figure: 7}. It is clear from
this figure that the set of parameters (SET 3) given in Table \ref{tab:condensate} and in Eq. (\ref{lambda2}) give a much better fit
of the data. 
\section{\boldmath$ \lambda^2$ and the light quark masses}
\nin
\b~In the (pseudo)scalar channels, $\lambda^2$ decreases the estimate of the light quark mass by about 5-6\%
\cite{CNZ}, implying the value to order $\alpha_s^3$ \cite{SNB,SNLH}:
\beq
 (\overline{m}_u+\overline{m}_d)(2 ~\rm{GeV})=(8.6\pm 2.1)~\rm{MeV}~.
\eeq
 Using the ChPT ratio:
$
{2m_s/ m_d+m_u}=24.4\pm 1.5
$\cite{GASSER}, one can deduce:
\beq\label{eq:mspion}
\overline{m}_s(2 ~\rm{GeV})=(105\pm 26)~\rm{MeV}~.
\eeq
\b~For the extraction of the strange quark mass from ~$\tau$-decay, one works 
with a suitable combination of finite energy sum rules for the $\Delta S=-1$ component of $\tau$-decay \cite{SNMS}:
\bea {\cal S}_{10}&\equiv& \int_0^{t_c} dt \ga 1-2{t\over t_c}\dr
\frac{1}{\pi}{\rm Im}\Pi^{(0+1)}_{V+A}~,
\eea
which is directly sensitive to~ $m_s^2$ and to $\lambda^2$ to leading order and which has a convergent PT series
(optimal $t_c\approx M^2_\tau$).
One finds that $m_s$ increases with $|\lambda^2|$. For the value of $\lambda^2$ given in Eq. (\ref{lambda2}), one
obtains:
\bea\label{eq:mstau}
\hat m_s&=&(106^{+33}_{-37})~{\rm MeV}\lrar\nnb\\ &&\overline m_s(2~\rm{GeV})=(93^{+29}_{-32})~{\rm MeV}
\eea
As one can see in Table \ref{tab:ms}, the value of $\lambda^2=0$ which leads to a too small value of $m_s$
is excluded by the lower bound from (pseudo)scalar channels \cite{LELLOUCH} updated
to order $\alpha_s^3$ and including $\lambda^2$ to be ($71.4\pm 3.7$) MeV in \cite{SNB,SNLH}, and by the bound of about 80 MeV from the
direct extractions of the quark condensate \cite{DOSCH2}.
\vspace{-1cm}
\begin{center}
\begin{table}[hbt]
\setlength{\tabcolsep}{0.pc}
\caption{Invariant mass $\hat m_s$ from the tau-decays}
\label{tab:ms}
\begin{tabular*}{7cm}{@{}l@{\extracolsep{\fill}}cc}
\hline
$-a_s\lambda^2$ in GeV$^2$&$\hat m_s$ in MeV\\
\hline
0.03& $56\pm 28\pm 8$\\
0.06&$81\pm 21\pm 9$\\
0.07&$89\pm 21\pm 9$\\
0.09&$102\pm 18\pm 11$\\
0.12&$118\pm 16\pm 11$\\
0.15&$132\pm 15\pm 11$\\
0.18&$145\pm 13\pm 12$\\
\hline
\end{tabular*}
\end{table}
\end{center}
\vspace{-1.2cm}
\nin
Combining the previous values in Eqs. (\ref{eq:mspion}) and (\ref{eq:mstau}) with the value:
\beq\overline{m}_s(2
~\rm{GeV}) =(81\pm 22)~{\rm MeV}~,
\eeq
obtained from tau-decays using the difference of the $\Delta S=-1$ with
the $\Delta S=0$ spectral functions \cite{PICH}\footnote{The inclusion of
$\lambda^2$ increase the original result by $0.4\%$ which is negligible.}, one can deduce the average\footnote{We have not included the result from $e^+e^-\rar$ hadrons data \cite{SNE}, which we are revisiting.}:
\beq\la\overline{m}_s(2
~\rm{GeV})\ra =(92\pm 15)~{\rm MeV}~,
\eeq
which can be compared with lattice results \cite{LATT}. 
\section{\boldmath $\lambda^2$ and the value of $\alpha_s$ from tau decay}
\nin
Including $\lambda^2$, the   modified expression    of the tau hadronic width is \cite{BNP}:
\bea
     R_\tau&=& 3\left( (|V_{ud}|^2+|V_{us}|^2\right) S_{EW}^\tau  \times\nnb\\
&&\Big{\{}
1+\delta^\tau_{EW}+\delta^\tau_{PT}+ \delta^\tau_2+
\delta^\tau_{\lambda}+\delta^\tau_{NP}\Big{\}}.\eea    
{where:        $|V_{ud}|=0.9751\pm 0.0006$ and $|V_{us}| =0.221\pm 0.003$ are the weak mixing
angles; $S^\tau_{EW}=1.0194$ and $~~~\delta^\tau_{EW} =0.0010$ are the electroweak
corrections; $\delta^\tau_{PT}=
a_s+5.2023a_s^2+26.366a_s^3+...$ is the pQCD series using Fixed Order Perturbation Theories (FOPT);
... are higher order uncalculated terms which, at present, are model dependent. In \cite{BNP,PDG}, it has been taken to be
$\approx \pm 130 a_s^4
\simeq
\pm 0.020$, which is the main source of theoretical error. 
     In the present approach, this term is replaced by the tachyonic gluon correction \cite{CNZ}:
 \beq
\delta^\tau_\lambda\simeq  -2 \times 1.05{a_s\lambda^2\over{M^2_\tau}}\simeq 0.0465\pm 0.0199.
\eeq
which has about the same size as the $\alpha_s^3$ correction! One can notice that one does not gain much on the
precision but the origin of the error is more justified and can be improved; $\delta_2^\tau$ is the tiny running light quark
masses corrections and $\delta^\tau_{NP}=-(2.8\pm 0.6) \times 10^{-2}$ is the sum of non-perturbative effects of operators
of dimension D $\geq$ 4} \cite{SNG}. Using the average of the hadronic width from the tau-lifetime and leptonic widths:
\beq
\la R_\tau^{exp}\ra= 3.634\pm 0.004~,
\eeq
one can deduce:
$\delta^\tau_{PT}\simeq 0.1688\pm 0.0199$. Then:
\bea \alpha_s(M_\tau)=&&0.325\pm 0.025   \rm  (Contour Coupling)\cite{DIBERDER}\nnb\\
&& 0.303\pm 0.022~  \rm (FOPT)\cite{BNP}~,
\eea
giving the average:
\beq 
\alpha_s(M_\tau) = 0.313\pm 0.017~,
\eeq
which, runned until $M_Z$ becomes:
\beq 
\alpha_s(M_Z)= 0.117\pm 0.002~.
\eeq
Compared to previous estimates \cite{BNP,DIBERDER,PDG,ALEPH}, the presence of $\lambda^2$ shifts the central value of
$\alpha_s(M_Z)$ from 0.121 to 0.117, while the source of the error is understood and is slightly improved here.
This number agrees quite well with the existing world average $0.1187\pm 0.0020$ given by PDG \cite{PDG}.
\section{Conclusions}
\nin
We have shortly reviewed the different estimates of power corrections entering into the SVZ-expansion, and discussed
the phenomenology implied by the existence of a tachyonic gluon mass $\lambda^2$, which induces a new $1/Q^2$ term in the
OPE. We found that the size of higher dimension condensates deviates notably from the vacuum saturation assumption valid
at large $N_c$, while $\lambda^2$ improves the low-energy phenomenology. We plan to explore its effect in the heavy
quark sector \cite{CNZ2}.  
\section*{Acknowledgements}
\nin
Discussions with Valya Zakharov are gratefully acknowledged.

\end{document}